\documentclass[10pt,twocolumn,showpacs,aps,floatfix]{revtex4}
\usepackage{epsfig}
\usepackage{amsmath}
\usepackage{dcolumn}% Align table columns on decimal point
\usepackage{bm}% bold math
\usepackage{graphicx}
\usepackage{wrapfig}
\usepackage{graphicx}
\usepackage{epstopdf}
\usepackage{epsfig}
\usepackage{color}
\usepackage{IEEEtrantools}

\newcommand{\G}{\Gamma}

\newcommand{\g}{\gamma}
\newcommand{\dl}{\delta}
\newcommand{\ve}{\varepsilon}

\begin{document}

%------------------------------------------------------------

\title{Resonance width distribution for open quantum systems}

\author{Gavriil Shchedrin}
\email{http://www.nscl.msu.edu/~shchedri/}
\author{Vladimir Zelevinsky}
\email{http://www.nscl.msu.edu/~zelevins}

\affiliation{Department of Physics and Astronomy and
National Superconducting Cyclotron Laboratory,
Michigan State University, East Lansing, MI 48824, USA}

%------------------------------------------------------------

\begin{abstract}
Recent measurements of resonance widths for low-energy neutron scattering off heavy nuclei show large
deviations from the Porter-Thomas distribution.
We propose a ``standard'' width distribution based on the random matrix theory for a chaotic quantum system
with a single open decay channel.
Two methods of derivation lead to a single analytical expression that recovers,
in the limit of very weak continuum coupling,
the Porter-Thomas distribution.
The parameter defining the result is the ratio of typical widths $\Gamma$ to
the energy level spacing $D$. Compared to the Porter-Thomas distribution, the new distribution
suppresses small widths and increases the probabilities of larger widths. We show also that it is necessary
to take into account the gamma channels.
\end{abstract}

%------------------------------------------------------------


\maketitle

%------------------------------------------------------------

\section{Introduction}
Random matrix theory as a statistical approach for exploring properties of complex quantum
systems was pioneered by Wigner and Dyson half a century ago \cite{Mehta}. This theory
was successfully applied to excited states of complex nuclei and other mesoscopic systems \cite{Bohr1,big,stoeckmann00,WMRMP09}, evaluating statistical fluctuations and correlations
of energy levels and corresponding wave functions supposedly of ``chaotic" nature.

The standard random matrix approach based on the Gaussian Orthogonal Ensemble (GOE) for systems
with time-reversal invariance,
and on the Gaussian Unitary Ensemble (GUE) if this invariance is violated,
was formulated originally for {\sl closed} systems with no
coupling to the outside world. Although the practical studies of complex nuclei, atoms, disordered
solids, or microwave cavities always require the use of reactions produced by external sources,
the typical assumption was that such a probe at the resonance is sensitive to the specific components
of the exceedingly complicated intrinsic wave function, one for each open reaction channel, and
the resonance widths are measuring the weights of these components \cite{brody81}. With the Gaussian
distribution of independent amplitudes in a chaotic intrinsic wave function, the widths under this
assumption are proportional to the squares of the amplitudes and as such can be described, for
$\nu$ independent open channels, by the chi-square distribution with $\nu$ degrees of freedom.
For low-energy elastic scattering of neutrons off heavy nuclei, where the interactions can be considered
time-reversal invariant, one expects $\nu=1$ that is usually
called the Porter-Thomas distribution (PTD) \cite{PT56}.

\begin{figure}\label{fig1}
\centerline{\mbox{\epsfxsize=8.2cm \epsffile{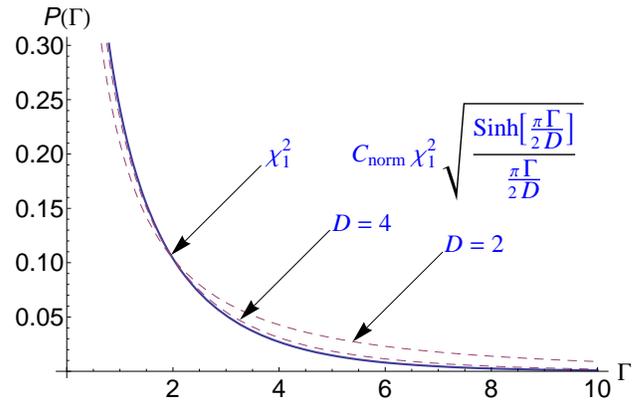}} } \caption{
The proposed resonance width distribution
according to eq. (\ref{2}) with a single neutron channel
in the practically important case $\eta\gg \Gamma$. The width $\Gamma$
and mean level spacing $D$ are measured in units of the mean value $\langle\Gamma\rangle$.}
\end{figure}

\begin{figure}[b!]\label{fig2}
\centerline{\mbox{\epsfxsize=8.2cm \epsffile{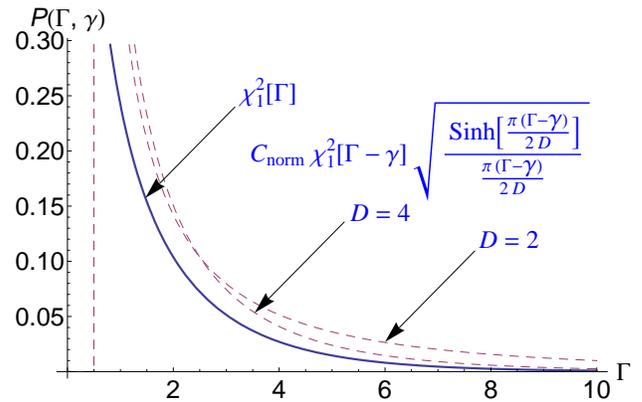}} } \caption{
The proposed resonance width distribution
according to eq. (\ref{27}) with a single neutron channel and $N$ gamma-channels
in the practically important case $\eta\gg \Gamma$. The neutron width $\Gamma$, radiation
width $\gamma$, and
mean level spacing $D$ are measured in units of the mean value $\langle\Gamma\rangle$.}
\end{figure}

Recent measurements \cite{koehler10,koehler11} claimed that the neutron width distributions in low-energy
neutron resonances on certain heavy nuclei are different from the PTD. As a rule, the fraction of
greater widths is increased, while the fraction of narrow resonances is reduced which, being
approximately presented with the aid of the same standard class of functions, would require $\nu\neq 1$.
The literature discussing the scattering and decay processes in chaotic systems, see for example
\cite{FS2,KH1,FS1,SS1} and references therein, does not provide a detailed description of the
width distribution for the region of relatively small widths as observed in low-energy
neutron resonances.

There are various reasons for possible deviations from the simple statistical predictions
\cite{weiden10,celardo11,volya11}. First of all, the intrinsic dynamics, even in heavy nuclei, can be
different from that in the GOE limit of many-body quantum chaos. If so, the detailed analysis of
specific nuclei is required. As an example we can mention $^{232}$Th, where for a long time
a {\sl sign problem} exists \cite{mitchell01} concerning the resonances with strong enhancement of parity
non-conservation in scattering of longitudinally polarized neutrons. The observed predominance of a certain
sign of parity violating asymmetry contradicts to the statistical mechanism of the effect and may
be related to the non-random coupling between quadrupole and octupole degrees of freedom \cite{FZ95}.
The width distribution in the same nucleus reveals noticeable deviations from the PTD. The presence
of a shell-model single-particle resonance serving as a doorway to the compound nucleus can also make
its footprint distorting the statistical pattern. Another (maybe related to the doorway resonance)
effect can come from the changed energy dependence of the widths that is usually assumed to be
proportional to $E^{\ell+1/2}$ for neutrons with orbital momentum $\ell$. Finally, the situation
is not strictly one-channel, since, along with elastic neutron scattering, many gamma-channels are
open as well. However, apart from structural effects, even in one-channel approximation, there exists
a generic cause for the deviations from the PTD, since the applicability of the GOE is anyway violated
by the {\sl open} character of the system \cite{SZ89}. The appropriate modification of the GOE and PTD
predictions, which should be applied before making specific conclusions, is our goal below.

The resonances are not the eigenstates of a Hermitian Hamiltonian, they are poles of the scattering
matrix in the complex plane. Their {\sl complex energies} $\mathcal{E}=E-i{\G}/{2}$ can be rigorously described
as eigenvalues of the {\sl effective non-Hermitian Hamiltonian} \cite{MW}. As shown long ago, even for
a single open channel, the statistical properties of the complex energies cannot be described by the GOE.
The new dynamics is related to the interaction of intrinsic states through continuum. In the limit of
strong coupling this leads to the overlapping resonances, Ericson fluctuations of cross sections, and
sharp redistribution of widths similar to the phenomenon of {\sl super-radiance}, see the review \cite{AZ11} and
references therein. The control parameter of such restructuring is the ratio
\begin{IEEEeqnarray}{lll}                                                   \label{1}
\kappa=\frac{\pi\Gamma}{2D}
\end{IEEEeqnarray}
of typical widths, $\Gamma$, to the mean spacing between the resonances, $D$. In the region of low-energy
neutron resonances, $\kappa$ is still small but in order to correctly separate the general statistical
effects from peculiar properties of individual nuclei we need to have at our disposal a generic width
distribution that differs from the PTD as a function of the degree of openness.

\section{Resonance width distribution}

The goal of the paper is to provide a practical tool that would allow one to compare an experimental
output for an unstable quantum system with predictions of random matrix theory. We propose a new distribution
function that is based, similar to the GOE, on the chaotic character of time-reversal invariant internal dynamics
and corresponding decay amplitudes, but properly accounts for the continuum coupling through the effective non-Hermitian Hamiltonian. The numerical simulations for this Hamiltonian were described earlier \cite{mizutori93,celardo11} but here we derive the analytical expression. We limit ourselves here by
the situation typical for nuclear applications, namely $\kappa<1$. The super-radiant regime, $\kappa\geq 1$,
can be of special interest, including such systems as microwave cavities, and in the considered framework
the formal symmetry exists, $\kappa\rightarrow 1/\kappa$. At a large number of resonances and fixed number
of open channels, after the super-radiant transition the broad state becomes a part of the background while the remaining ``trapped" states return into the non-overlap regime. However, in heavy nuclei this transition hardly
can be observed because earlier many new channels can be opened; in the modification of the PTD we see only
precursors of this transition.

Our arguments will follow two different routes which lead to the equivalent results. The final formula
for the statistical width distribution can be presented as
\begin{IEEEeqnarray}{lll}                                                   \label{2}
P(\Gamma)=C
\frac{\exp{\left[-\frac{N}{2\sigma^{2}}
\Gamma(\eta-\G)\right]}}
{\sqrt{\Gamma(\eta-\Gamma)}}
\left(\frac{\sinh{\left[\frac{\pi \Gamma}{2D}\frac{(\eta-\Gamma)}{\eta}\right]}}
{\frac{\pi \Gamma}{2D}\frac{(\eta-\Gamma)}{\eta}}\right)^{1/2}.
\end{IEEEeqnarray}
Here we consider $N\gg 1$ intrinsic states coupled to a single decay channel, for example, $s$-wave elastic
neutron scattering. The parameter $D$ is a mean energy spacing between the resonances,
$\kappa$ is a new dimensionless combination, eq. (\ref{1}), and $C$ is a normalization constant.
The quantity $\eta$ is the total sum (the trace of the imaginary part of the effective non-Hermitian Hamiltonian
that remains invariant in the transition to the biorthogonal set of its eigenfunctions)
of all $N$ widths; it appears as a parameter that fixes the starting ensemble distribution, see eqs. (\ref{8})
and (\ref{9}). The possible values of widths are restricted from both sides, $0<\Gamma<\eta$. The above mentioned symmetry $\kappa\rightarrow 1/\kappa$ is reflected in the symmetry
$\Gamma\rightarrow \eta-\Gamma$ of a factor in eq. (\ref{2}) but, as was already stated, our region of interest is
at $\Gamma\ll \eta$. Another parameter, $\sigma$, determines the standard deviation of
variable $\G$ evaluated consistently with the distribution of eq. (\ref{2}).
In the practical region far away from the super-radiance we obtain
\begin{IEEEeqnarray}{lll}                                                   \label{3}
P(\Gamma)=C \chi^{2}_{1}(\Gamma)
\left(\frac{\sinh{\kappa}}{\kappa}\right)^{1/2}.
\end{IEEEeqnarray}
The PTD is recovered in the limiting case $\kappa\ll 1$ that corresponds to the approximation
of an open quantum system by a closed one. The new element is the factor explicitly determined by the coupling
strength $\kappa$. With growing continuum coupling the probability of larger widths
increases.  The distribution (\ref{2}) for different ratios $\langle\Gamma\rangle/D$ is shown in Fig. 1.

The origin of the square root in the new factor is the linear energy level repulsion typical for the GOE
spectral statistics. Indeed, in the complex plane, $\mathcal{E}=E-i{\G}/{2}$, the distance between two poles $\mathcal{E}_{m}$ and $\mathcal{E}_{n}$ is  $\sqrt{(E_{m}-E_{n})^2+{(\Gamma_{m}-\Gamma_{n})^2}/{4}}$; after
integration over all variables of other states we obtain a characteristic square root in the
level repulsion, see below eq. (\ref{12}) and the discussion after eq. (\ref{21}).

\section{Effective non-Hermitian Hamiltonian and scattering matrix}

In order to come to the result (\ref{2}), we start with the general description of complex
energies $\mathcal{E}=E-i{\G}/{2}$ in a system of $N$ unstable states satisfying the GOE statistics inside the
system and interacting with the single open channel through Gaussian random amplitudes.
The general reaction theory \cite{LT1} is constructed in terms of the elements of the scattering matrix in
the space of open channels $a,b,...$,
\begin{IEEEeqnarray}{lll}                                                   \label{4}
S^{ba}(E)=\dl^{ba}-iT^{ba}(E).
\end{IEEEeqnarray}
Within the formalism of the effective non-Hermitian Hamiltonian ${\cal H}=H-(i/2)W$, the $T$-matrix is defined as
\begin{IEEEeqnarray}{lll}                                                   \label{5}
T^{ba}(E)=\sum_{m,n=1}^{N}A_{m}^{b{\ast}}\left[\frac{1}{E-\mathcal{H}}\right]_{mn}A_{n}^{a}
\end{IEEEeqnarray}
in terms of the amplitudes $A_{n}^{a}$ connecting an internal basis state $n$ with an open channel $a$.
Here we neglect the potential part of scattering that is not related to the internal dynamics of
the compound nucleus.
The anti-Hermitian part of the effective Hamiltonian is exactly represented by the sum over $k$ open channels,
\begin{IEEEeqnarray}{lll}                                                   \label{6}
W_{mn}=\sum_{a=1}^{k}A_{m}^{a}A_{n}^{a{\ast}},
\end{IEEEeqnarray}
where the amplitudes can be considered real in the case of time-reversal invariance. It is important
that the factorized structure of the effective Hamiltonian guarantees the unitarity of the scattering
matrix. The amplitudes $A^{a}_{n}$ are uncorrelated Gaussian quantities with zero mean and variance
defined as $\overline{A^{a}_{n}A^{b}_{n'}}=\delta^{ab}\delta_{nn'}\eta/N$. The trace of the anti-Hermitian
part of the effective Hamiltonian, $\eta={\rm Tr}W$, i.e. the total sum of all $N$ widths used in eq. (\ref{2}),
is a quantity invariant under orthogonal transformation of the intrinsic basis. The detailed discussion of the whole approach, numerous applications and relevant references
can be found in the recent review article \cite{AZ11}.

The simplest version of the $R$-matrix description uses instead of the amplitude $T^{ba}$ its approximate
form, where the denominator contains poles on the real energy axis corresponding to the eigenvalues
of the Hermitian part $H$ of the effective Hamiltonian. Then the continuum coupling occurs only at the entrance
and exit points of the process while the influence of this coupling on the intrinsic dynamics of the compound
nucleus is neglected (in general, $H$ should also be renormalized by the off-shell contributions from the presence of the decay channels). Contrary to that, the full amplitude $T^{ba}$, eq. (\ref{5}), accounts for this coupling during the entire process including the virtual excursions to the continuum and back from intrinsic states. The poles
are the eigenvalues of the full effective Hamiltonian in the lower half of the complex energy plane.
The experimental treatment corresponds to this full picture.
According to the original paper \cite{koehler10}, the $R$-matrix code SAMMY \cite{Lar1} had been used
in the experimental analysis where the relevant expression is given in the form
\begin{IEEEeqnarray}{lll}                                                   \label{7}
R_{cc'}=\sum_{\lambda}\frac{\g_{\lambda{c}} \g_{\lambda{c'}} }
{E_{\lambda}-E-i \Gamma_{\lambda}/2}\dl_{JJ'},
\end{IEEEeqnarray}
and the treatment included a careful segregation of $s$- and $p$-resonances, $J=J'=1/2$ for an even target nucleus.
In the notations of \cite{Lar1} $\lambda$ represents a particular resonance, $E_{\lambda}$ is the energy of the resonance. Here we can identify the intermediate states $\lambda$ and
their complex energies $E_{\lambda}-i\Gamma_{\lambda}/2$ with the eigenstates and complex eigenvalues of ${\cal H}$,
while the numerator includes the amplitudes transformed to this new basis (under time-reversal invariance
the scattering matrix is symmetric). In terms of the reduced width $\gamma^{2}_{\lambda {c}}$ and the penetration factor $P_{c}$, the partial width is $\Gamma_{\lambda {c}}=2P_{c}\gamma^{2}_{\lambda {c}}$. Assuming a single channel
and universal energy dependence of penetration factors, the statistics of the total widths is the same as
that of $\gamma^{2}_{\lambda c}$.

\section{From ensemble distribution to single width distribution}

For a single-channel case, the {\sl joint distribution} ${P}(\vec{E};\vec{\Gamma})$ of all complex energy poles
has been rigorously derived in \cite{SZ89} under assumptions of the GOE intrinsic dynamics in the closed system
and Gaussian distributed random decay amplitudes. The result is given by
\begin{widetext}
\begin{IEEEeqnarray}{lll}                                                   \label{8}
{P}(\vec{E};\vec{\Gamma})=
C_{N}
\prod_{m<n}\frac{(E_{m}-E_{n})^2+\frac{(\Gamma_{m}-\Gamma_{n})^2}{4}}
{\sqrt{(E_{m}-E_{n})^2+\frac{(\Gamma_{m}+\Gamma_{n})^2}{4}}}
\prod_{n}\frac{1}{\sqrt{\Gamma_{n}}}\,
e^{-NF(\vec{E};\vec{\Gamma})},
\end{IEEEeqnarray}
\end{widetext}
where the ``free energy" $F$ contains interactions of $N\gg 1$ complex poles in the interval $2a=ND$ of energies,
\begin{IEEEeqnarray}{lll}                                                   \label{9}
F(\vec{E};\vec{\Gamma})=\frac{1}{a^2}\sum_{n}E_{n}^2+\frac{1}{2a^2}
\sum_{m<n}\Gamma_{m}\Gamma_{n}+\frac{1}{2\eta}\sum_{n}\Gamma_{n}.
\end{IEEEeqnarray}
For given $N$, this distribution contains two parameters, the semicircle radius $a$ for the intrinsic dynamics
and $\eta$ characterizing the total trace of the imaginary part of the effective Hamiltonian.

Considering this free energy in the ``mean-field" approximation, we see that the original mean value $\langle\Gamma\rangle_{0}=\eta/N$ is substituted by $\langle\Gamma\rangle$ that is determined by
the competition of two terms, $1/\langle\Gamma\rangle=1/\langle\Gamma\rangle_{0}+\langle\Gamma\rangle/4D^{2}$.
The first product in front of $\exp(-NF)$ in eq. (\ref{8}) substitutes the GOE level repulsion by the repulsion
in the complex plane and interaction of the poles with their negative-$\Gamma$ ``images" \cite{SZ89}.
The structure of this result guarantees that all widths $\Gamma$ are positive. The difficulty with the
distribution of eq. (\ref{8}) is that it is {\it not an analytic function} of complex energies.

Our first step is to specify a single $N$-th pole $(E_{N},\Gamma_{N})\equiv{}(E,\Gamma)$ and, using the fact that the
distribution ensures $\Gamma_{n}\geq{0}$, return to the absolute values of the amplitudes,
$\sqrt{\Gamma_{n}}=\xi_{n}$ for other roots. In this form we can apply the steepest descent
method owing to a large parameter $N\gg1$ and a {\sl saddle point} inside the integration
interval that was absent in the initial expression. Integration $\prod_{n=1}^{N-1}{d{\xi_{n}}}$ over all
but one variable $\Gamma\equiv{}\Gamma_{N}=\xi_{N}^{2}$ leads to the following result for ${P}(\vec{E};\Gamma)$ as a function of multiple energy variables $\vec{E}$
and a single width variable $\Gamma$:
\begin{widetext}
\begin{IEEEeqnarray}{lll}                                                   \label{10}
{P}(\vec{E};\Gamma)=
C_{N}\prod_{m<n}
\left|E_{m}-E_{n}\right|
\prod_{n}
{\sqrt{(E_{n}-E)^2+\frac{\G^2}{4}}}
\exp\left[
-\frac{N}{a^{2}}\left(\sum_{n}{E_{n}^2}+E^2
\right)\right]
\frac{\exp\left[
-\frac{N}{2\eta}\G\right]}{\sqrt{\G}}
\left(\sqrt{\frac{2\pi}{N(\frac{\G}{a^2}+\frac{1}{\eta})}}\right)^{N-1}
\end{IEEEeqnarray}
\end{widetext}
Introducing new variables, $2a/N=D$ and $\lambda=\eta N$, we shall examine the behavior of
one of the $N$-dependent factors in eq. (\ref{10}) in the limit of $N\rightarrow\infty$:
\begin{widetext}
\begin{IEEEeqnarray}{lll}                                                   \label{11}
\lim_{N\rightarrow{\infty}}\left[\frac{\exp\left[
-\frac{N}{2\eta}\G\right]}{\sqrt{\G}}
\left(1+\frac{\lambda\G}{a^2N}\right)^{-\frac{N-1}{2}}\right]=\frac{\exp\left[
-\frac{N}{2\eta}\G\right]}{\sqrt{\G}}
\exp\left[-\frac{\lambda}{2a^2}\G\right]
\end{IEEEeqnarray}
\end{widetext}
The scaling properties of the parameters $\eta$, $a^{2}$, and $\lambda$
are as follows: $\eta\propto{N}$, $a^{2}$ and $\lambda$ are both $\propto N^{2}$.
As a result, the product of two exponents produces a well defined
limit that brings in the desired dependence on the coupling strength $\kappa$.

The real energy distribution does not change much in an open system with a single decay channel being
still, at finite but large $N$, close to a semicircle.
We are working in the central region of the spectrum where the level
density is approximately constant and the energy spectrum is close to equidistant (the maximum of
the level spacing distribution is always at $s=\delta E/D\approx 1$ although the distribution in an open system
changes at small spacings, $s<1$, as we will comment later). With $E_{n}=E+nD$, we are able to perform
an exact calculation of the product:
\begin{widetext}
\begin{IEEEeqnarray}{lll}                                                   \label{12}
C_{N}
\prod_{n}{\sqrt{(E_{n}-E)^2+\frac{\G^2}{4}}}=\widehat{C}_{N}
\left(\prod_{n=1}^{N}\left[1+\frac{\G^{2}/4}{(nD)^{2}}\right]\right)^{1/2}_{N\rightarrow{\infty}}=
\widehat{C}_{N}
\left(\frac{\sinh\left[\frac{\pi}{2}\frac{\G}{D}\right]}{\frac{\pi}{2}\frac{\G}{D}}\right)^{1/2},
\end{IEEEeqnarray}
\end{widetext}
where we have used the famous {\sl Euler formula},
\begin{IEEEeqnarray}{lll}                                                   \label{13}
\frac{\sinh{x}}{x}=\prod_{k=1}^{\infty}\left[1+\frac{x^{2}}{k^{2}\pi^{2}}\right].
\end{IEEEeqnarray}
The width-independent factors will enter the normalization constant.
Of course, the whole reasoning is valid in the limit $N\gg 1$.
Finally, the  width distribution for $\Gamma\ll \eta$ is represented by
\begin{widetext}
\begin{IEEEeqnarray}{lll}                                                   \label{14}
{P}(\Gamma)=
C\left(\frac{\sinh\left[\frac{\pi}{2}\frac{\Gamma}{D}\right]}{\frac{\pi}{2}\frac{\Gamma}{D}}\right)^{1/2}
\frac{\exp\left[-\frac{N}{2\eta}\G\right]}{\sqrt{\G}}
\exp\left[-\frac{\lambda}{2a^2}\G\right].
\end{IEEEeqnarray}
\end{widetext}

\section{Doorway approach}

As an alternative derivation, we will apply the {\sl doorway approach} \cite{Bohr1,SZ92,AZ07}. Here we use
the eigenbasis of the imaginary part $W$ of the effective non-Hermitian Hamiltonian. Due to the factorized
nature of $W$  dictated by unitarity \cite{SZ89}, the number of its non-zero eigenvalues is equal to the number
of open channels. In our case we have only one eigenvalue, the {\sl doorway} $\varepsilon_{0}-i\eta/2$, that
has a non-zero width equal to the imaginary part $\eta$ of the trace of the Hamiltonian. Remaining basis
states are stable being driven by the Hermitian intrinsic Hamiltonian; its diagonalization produces
their real energies $\varepsilon_{n}$. These states acquire the widths through the interaction with
the doorway state; the corresponding matrix elements will be denoted $h_{n}$. In this basis, the Hamiltonian is represented as
\begin{IEEEeqnarray}{lll}                                                   \label{15}
\left(
  \begin{array}{ccccc}
    \varepsilon_{0} - \frac{i}{2}\eta & h_{1} & h_{2} & \cdots & h_{N} \\
    h^{*}_{1} & \varepsilon_{1} & 0 & \cdots & 0 \\
    h^{*}_{2} & 0 & \varepsilon_{2} & \cdots & 0 \\
    \cdots & \cdots & \cdots & \cdots & \cdots \\
    h^{*}_{N} & 0 & 0 & \cdots & \varepsilon_{N} \\
  \end{array}
\right)
\end{IEEEeqnarray}

The complex eigenvalues $\mathcal{E}=E-i{\G}/{2}$ are the roots of the secular equation,
\begin{IEEEeqnarray}{lll}                                                   \label{16}
\mathcal{E}=\varepsilon_{0}-\frac{i}{2}\eta+\sum_{n=1}^{N}\frac{|h_{n}|^{2}}{\mathcal{E}-\varepsilon_{n}},
\end{IEEEeqnarray}
that is equivalent to the set of coupled equations for real and imaginary parts,
\begin{IEEEeqnarray}{lll}                                                   \label{17}
E=\varepsilon_{0}+\sum_{n=1}^{N}\left|h_{n}\right|^{2}\frac{E-\varepsilon_{n}}{(E-\varepsilon_{n})^{2}+
\Gamma^{2}/4},
\end{IEEEeqnarray}\\
\\
\begin{IEEEeqnarray}{lll}                                                   \label{18}
\G=\frac{\eta}{1+{\sum_{n=1}^{N}\frac{\left|h_{n}\right|^{2}}{(E-\ve_{n})^{2}+\G^{2}/4}}}\equiv
f(\Gamma,E).
\end{IEEEeqnarray}\\
\\
For the Gaussian distribution of the coupling matrix elements with
$\langle \hspace{1mm} |h|^{2} \hspace{0.5 mm} \rangle=2\sigma^{2}/N$
(this scaling was derived in \cite{SZ92}), we obtain

\begin{widetext}
\begin{IEEEeqnarray}{lll}                                                   \label{19}
{P}(\G)=\int_{-\infty}^{+\infty}
\delta\left(\Gamma-f(\Gamma,E)
\right)
\exp{\left[-\frac{N}{\sigma^{2}}\sum_{n=1}^{N}h_{n}^2\right]}
\prod_{n=1}^{N}d{h_{n}}.
\end{IEEEeqnarray}
\end{widetext}
The integration in (\ref{19}) via the steepest descent method leads to eq. (\ref{2}).
In order to get this result we use a possibility to find a highest root $E=\ve_{N}$
which we set as an origin relative to which the energies $\ve_{n}$
can be counted as $E=\ve_{N},\;\ve_{n}=\ve_{N}-nD$.
An important intermediate step is the evaluation of the infinite product
of the Lorentzian peaks that can be simplified as
\begin{widetext}
\begin{IEEEeqnarray}{lll}                                                   \label{20}
\left({\prod_{n=1}^{N-1}\left[1-\frac{\frac{\Gamma^{2}}{4}+(E-\varepsilon_{N})^{2}}
{\frac{\Gamma^{2}}{4}+(E-\varepsilon_{n})^{2}}\right]}\right)^{-1/2}=
\left({\prod_{n=1}^{N-1}\left[1-\frac{\frac{\Gamma^{2}}{4}}{\frac{\Gamma^{2}}{4}+(nD)^{2}}
\right]}\right)^{-1/2}=\left(\frac{\sinh\left[\frac{\pi}{2}\frac{\Gamma}{D}\right]}
{\frac{\pi}{2}\frac{\Gamma}{D}}\right)^{+1/2}.
\end{IEEEeqnarray}
\end{widetext}

In a similar way one can analyze the resonance spacing distribution $P(s)$ along the real energy axis;
spacings $s=\delta E/D$ are measured in units of their mean value $D$.
As predicted in \cite{SZ89} and observed numerically in \cite{mizutori93}, the short-range
repulsion disappears and the Wigner surmise with the standard linear preexponential factor $s$ is
substituted by the square root,
\begin{IEEEeqnarray}{lll}                                                   \label{21}
P(s)\propto \sqrt{s^{2}+4\frac{\langle \Gamma^{2}\rangle}{D^{2}}}
\exp\left[-\,{\rm const}\cdot s^{2}\right].
\end{IEEEeqnarray}
At spacing $s\ll 1$, the probability behaves as $a+bs^{2}$ with the quadratic dependence on $s$ that,
similar to the GUE, mimics the violation of time-reversal invariance due to
the open decay channel. The absence of short-range repulsion, $a\neq 0$ (the interaction through
continuum, opposite to a normal Hermitian perturbation, repells widths and attracts real energies
\cite{brentano96}), reflects the energy uncertainty of unstable states.

We demonstrated that two complementary approaches which reflect different physical aspects
of the situation lead essentially to the equivalent (after identification of corresponding
parameters) results which we prefer to write in the form (\ref{2}). We expect that for other
canonical ensembles the width distribution far from the super-radiance can be expressed by a similar
formula with the function $(\sinh\kappa/\kappa)^{\beta/2}$, where the standard index of ensemble
is $\beta=1$ for the GOE, $\beta=2$ for the GUE, and $\beta=4$ for the Gaussian Symplectic Ensemble.
In the same way we expect the square root in eq. (\ref{21}) to be substituted by the same
power $\beta/2$.

The doorway approach
naturally indicates the limits of the variable, $0\leq\Gamma\leq\eta$. It has also an advantage
of the possibility to generalize the answer taking into account explicitly the {\sl rigidity}
of the internal energy spectrum with fluctuations of level spacings around their mean value $D$
[in our approximation, only this average value enters eq. (\ref{2})]. Another direction of
generalization includes the possible influence of a single-particle resonance depending on
a position of its centroid with respect to the considered interval of the resonance spectrum. In particular,
that centroid may be located under threshold of our decay channel. In this case even
the standard energy dependence of the widths can change as was mentioned long ago \cite{SZ92},
see also \cite{weiden10}. The doorway state may or may not coincide with such a resonance so
that the effective Hamiltonian (\ref{15}) may contain two special states coupled with the ``chaotic"
background, one by intrinsic interactions and another one through the continuum.

\section{Adding gamma-channels}

The goal of this section is to estimate in the same spirit the influence of
$\gamma$-channels on the resonance width distribution. Only a single open elastic neutron channel was
taken into account in the analysis of data \cite{koehler10,koehler11}. The presence of even weak additional
open channels changes the unitarity conditions. Examples of mutual influence of neutron and gamma channels
are well known in the literature from long ago, see for example \cite{komano84}.

Generalizing the doorway description we allow now each intrinsic state to decay by gamma-emission
which is always possible independently of the position of the neutron threshold.
In the simplest approximation, the effective non-Hermitian Hamiltonian is now represented by
\begin{IEEEeqnarray}{lll}                                                   \label{22}
\left(
  \begin{array}{ccccc}
    \varepsilon_{0} - \frac{i}{2}\eta & h_{1} & h_{2} & \cdots & h_{N} \\
    h^{*}_{1} & \varepsilon_{1}-\frac{i}{2}\gamma_{1} & 0 & \cdots & 0 \\
    h^{*}_{2} & 0 & \varepsilon_{2}-\frac{i}{2}\gamma_{2} & \cdots & 0 \\
    \cdots & \cdots & \cdots & \cdots & \cdots \\
    h^{*}_{N} & 0 & 0 & \cdots & \varepsilon_{N}-\frac{i}{2}\gamma_{N} \\
  \end{array}
\right),
\end{IEEEeqnarray}
where we assumed that the intrinsic part of the matrix is pre-diagonalized and introduced $\gamma_{n}$ as
the widths for $\gamma$-channels.

Analogously to eq. (\ref{16}), the complex energy eigenvalues $\mathcal{E}=E-\frac{i}{2}\Gamma$
are the roots of the secular equation,
\begin{IEEEeqnarray}{lll}                                                   \label{23}
\mathcal{E}=\varepsilon_{0}-\frac{i}{2}\eta+\sum_{n=1}^{N}\frac{|h_{n}|^{2}}
{\mathcal{E}-\left(\varepsilon_{n}-\frac{i}{2}{\gamma_{n}}\right)},
\end{IEEEeqnarray}
equivalent to a set of coupled equations,
\begin{IEEEeqnarray}{lll}                                                   \label{24}
E=\varepsilon_{0}+\sum_{n=1}^{N}
\left|h_{n}\right|^{2}\frac{E-\varepsilon_{n}}{(E-\varepsilon_{n})^{2}+{(\G-{\gamma_{n}})^{2}}/{4}},
\end{IEEEeqnarray}

\begin{IEEEeqnarray}{lll}                                                   \label{25}
\G=\frac{\eta+\sum_{n=1}^{N}
\left|h_{n}\right|^{2}\frac{{\gamma_{n}}}{(E-\varepsilon_{n})^{2}+{(\G-{\gamma_{n}})^{2}}/{4}}}
{1+{\sum_{n=1}^{N}\left|h_{n}\right|^{2}\frac{1}{(E-\ve_{n})^{2}+{(\G-{\gamma_{n}})^{2}}/{4}}}}
\equiv g(\Gamma,E,\gamma).
\end{IEEEeqnarray}
The resonance width distribution for an open quantum system with $\gamma$-channels included is given by
\begin{widetext}
\begin{IEEEeqnarray}{lll}                                                   \label{26}
{P}(\G,\gamma)=\int_{-\infty}^{+\infty}
\delta\left(
\G-g(\Gamma,E,\gamma)
\right)
\exp{\left[-\frac{N}{\sigma^{2}}\sum_{n=1}^{N}h_{n}^2\right]}
\prod_{n=1}^{N}d{h_{n}}.
\end{IEEEeqnarray}
\end{widetext}

Estimating the gamma-widths by their average value, $\gamma$, and acting in the same manner as in the case
of a single open channel we come to
the final expression for the resonance width distribution,
\begin{widetext}
\begin{IEEEeqnarray}{lll}                                                   \label{27}                                                P(\Gamma,\gamma)=C
\frac{(\eta-\gamma)}{\sqrt{\Gamma-\gamma}\sqrt{\eta-\Gamma}}
\exp{\left[-\frac{N}{2\sigma^{2}}
(\Gamma-\gamma)(\eta-\G)\right]}
\left(\frac{\sinh\left[\frac{\pi (\Gamma-\gamma)}{2D} \frac{(\eta-\Gamma)}{\eta}\right]}
{\frac{\pi (\Gamma-\gamma)}{2D} \frac{(\eta-\Gamma)}{\eta}}\right)^{1/2},
\end{IEEEeqnarray}
\end{widetext}
that is shifted by $\Gamma\rightarrow\Gamma-\gamma$ compared to the previous result. The mentioned earlier symmetry
between the ends of the distribution, $\Gamma=0$ and $\Gamma=\eta$, would be substituted here by
$\Gamma \rightarrow {(\eta+\gamma)-\Gamma}$. Thus, the effective influence of $\gamma$-channels on the resonance width distribution is reduced here to a shift of the whole distribution by a mean radiation width $\gamma$ as seen in Fig. 2. In the practical region far away from the super-radiance, $\Gamma \ll {\eta}$, we obtain
\begin{widetext}
\begin{IEEEeqnarray}{lll}                                                   \label{28} 
P(\Gamma, \gamma)=\chi_{1}^{2}[\Gamma-\gamma]
\left(\frac{\sinh\left[\frac{\pi (\G-\g)}{2D}\right]}
{\frac{\pi (\G-\g)}{2D}}\right)^{1/2}.
\end{IEEEeqnarray}
\end{widetext}
In order to extract the neutron width from the total resonance width, the treatment of the data has to be modified making in a sense an inverse shift. Of course, a more precise consideration should use a statistical distribution
of the gamma widths.

\section{Conclusion}

In this article we propose a new resonance width distribution for an open quantum system based
on chaotic intrinsic dynamics and coupling of states with the same quantum numbers to the common
decay channel. Two approximate methods lead to an equivalent analytical expression for the width distribution
that does not belong to the class of chi-square distributions with the only parameter $\nu$
traditionally used in the analysis of data. In the limit of vanishing openness and return
to a closed system we recover the standard PTD. The new result
depends on the ratio (\ref{1}) of the width to the mean level spacing,
$\kappa\sim\Gamma/D$, that regulates the strength of the continuum coupling. The
deviations from the PTD grow with $\kappa$ up to the critical strength $\kappa\sim 1$, when
the broad ``super-radiant" state becomes essentially the part of the background, while the remaining
``trapped" states return to the weak coupling regime. This physics was repeatedly discussed
previously, especially in relation to quantum signal transmission through mesoscopic devices
\cite{AZ11,celardo10}, but it is outside of our interest here.

In the practical region of low-energy neutron resonances, the effects predicted here are relatively
small. Although at small $\kappa$ the derived neutron width distribution supports an experimental trend,
the final judgment can be made only after the presence
of gamma-channels was accounted for. We have to attract the attention of experimentalists to the fact that
the data should
be analyzed with the aid of the distribution that does not belong to the routinely used chi-square
class; gamma channels should be included into consideration. We can also mention that the result agrees
with numerical simulations \cite{celardo11}
for the full many-resonance distribution function (\ref{8}). Using the suggested distribution as a new
reference point, one can ascribe the remaining deviations to the specific features of individual systems
(level densities, single-particle structure in a given energy region, shape transformations,
energy dependence of the widths etc.).
Unfortunately, we still do not have experimental tests for the full distribution (\ref{8}).
Although in nuclear physics it is hard to make such a detailed analysis for higher energies and greater
degree of resonance overlap, the systems with tunable chaos, such as microwave cavities, acoustic
blocks, or even elastomechanical devices \cite{flores10}, seem to provide appropriate tools for such studies.

\section{Acknowledgements}

The authors thank P.E. Koehler and V.V. Sokolov for useful discussions, G.L. Celardo for numerical checks,
D.V. Savin for communication, M. Shapiro and S. Naboko for helpful remarks. G.S. is grateful to
the Guidance Committee members represented by N.O. Birge, B.A. Brown,
W.W. Repko, and R. Zegers for critical discussions. The authors are thankful to the Referee for a number of helpful suggestions and remarks.
Partial support from the NSF grant PHY-1068217 is gratefully acknowledged.

\end{document}